\documentclass[abstract=true]{scrartcl}

\usepackage{graphicx}
\usepackage{amssymb}
\usepackage{amsmath}
\usepackage{natbib}
\usepackage{subfig}

\usepackage[bookmarks,backref=false,linkcolor=black]{hyperref} 
\hypersetup{
  pdfauthor = {},
  pdftitle = {},
  pdfsubject = {},
  pdfkeywords = {},
  colorlinks=true,
  linkcolor= blue,
  citecolor= blue,
  pageanchor=true,
  urlcolor = blue,
  plainpages = false,
  linktocpage
}

\newcommand*{\doi}[1]{\href{https://dx.doi.org/\detokenize{#1}}{doi: \detokenize{#1}}}

\title{Orthogonal Edge Routing\\for the EditLens}

\author{
S. Gladisch$^1$ \and
V. Weigandt$^2$ \and
H. Schumann$^2$ \and
C. Tominski$^2$}
\date{}
\publishers{
$^1$Fraunhofer IGD, Rostock, Germany\\
$^2$Institute of Computer Science, University of Rostock, Germany
}

\begin{document}
\maketitle

\begin{abstract}

The EditLens is an interactive lens technique that supports the editing of graphs. The user can insert, update, or delete nodes and edges while maintaining an already existing layout of the graph. For the nodes and edges that are affected by an edit operation, the EditLens suggests suitable locations and routes, which the user can accept or adjust. For this purpose, the EditLens requires an efficient routing algorithm that can compute results at interactive framerates. Existing algorithms cannot fully satisfy the needs of the EditLens. This paper describes a novel algorithm that can compute orthogonal edge routes for incremental edit operations of graphs. Tests indicate that, in general, the algorithm is better than alternative solutions.

\end{abstract} 

\section{Introduction}
\label{routing}

The EditLens introduced by \cite{editlens} is an interactive tool for editing graphs. The EditLens enables users to insert, update, and delete nodes and edges of a graph while maintaining an already existing orthogonal layout of the graph. To relief users from time-consuming manual layout adjustments, the EditLens suggests suitable node positions and edge routes. Users can accept suggestions or request alternative suggestions by moving the EditLens across the existing graph layout. The EditLens has been developed in collaboration with domain experts from bio-informatics who need to maintain a constantly evolving molecular network, the so-called PluriNetwork~\cite{Som2010}.

To be operational, the EditLens requires an edge routing algorithm for a node-link representation that (i) computes orthogonal edge routes that are short, have a low number of edge bends, and circumvent existing nodes and (ii) has a runtime complexity that allows for interactive frame rates, so that users can explore different layout suggestions while moving the EditLens or switching between placement strategies (see \cite{editlens} for details).

\begin{figure}[t]
	\centering
	\includegraphics[width=\textwidth]{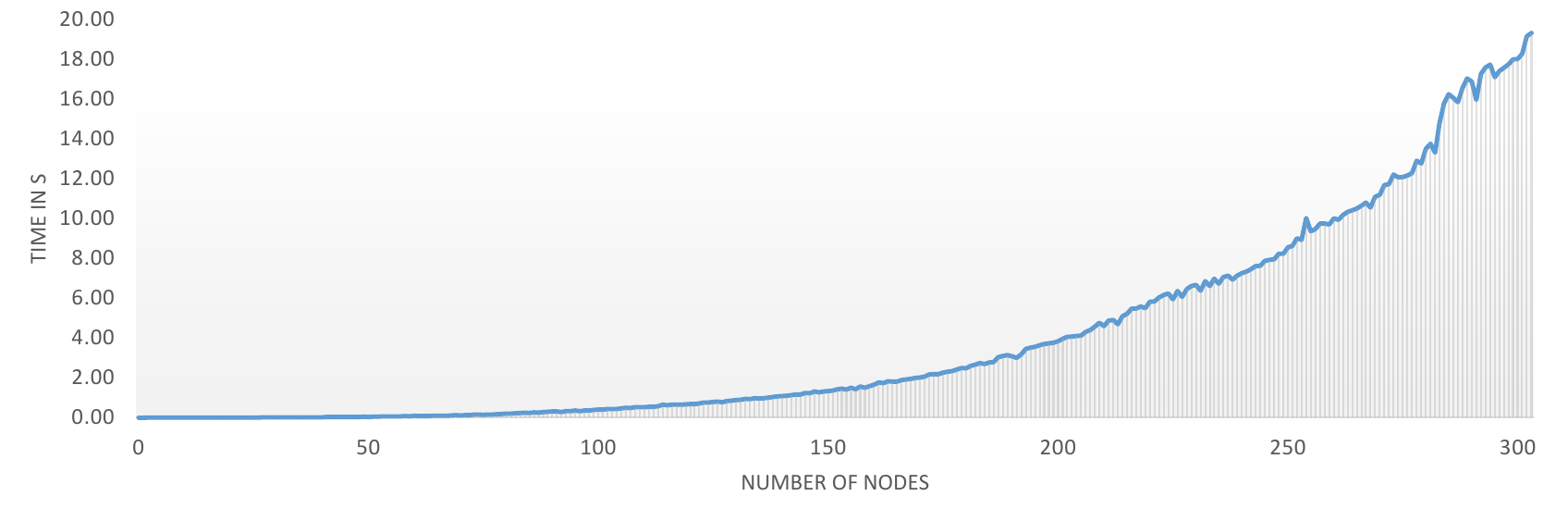}
	\caption[Edge routing runtime]{Results of a runtime performance test for computing orthogonal edge routes of a node-link layout with the edge routing algorithm by~\cite{Wybrow2009}.}
		\label{fig:chart}
\end{figure}

Wybrow et al.'s~\cite{Wybrow2009} algorithm has been identified as a promising candidate. It should be able to ``[...] calculate routings fast enough [...] during interaction.'' In order to confirm this, an experiment has been set up to test the algorithm's suitability for the EditLens.
A graph has been created automatically by incrementally inserting new nodes with edges to two randomly selected existing nodes. Each new node was placed on a randomly chosen free position in the node-link layout. On each insert, the edge routing for the graph was computed using the original implementation, and the runtime was measured.
Figure \ref{fig:chart} shows the average results of multiple test runs. The time needed for computing the edge routes increases rapidly as the graph grows. The computation of two edge routes in a node-link representation with 65 nodes and 130 edges took 100 ms on the test computer (Intel Core i7-3770k at 4.66 GHz, 16 GB RAM), which is already the upper limit for interactive frame rates. For 300 nodes and 600 edges, which is about the size of the PluriNetwork, the algorithm took nearly 20 seconds.
Therefore, a new orthogonal edge routing algorithm has been developed based on the algorithm by \cite{Wybrow2009}.

For the new algorithm it is assumed that nodes are represented with overlap-free circles whose quadratic bounding boxes leave a certain distance between each other.
The algorithm consists of the following three steps which are described in detail below.
\begin{enumerate}
\item Computation of potential edge routes
\item Search for suitable edge routes
\item Resolution of edge overlaps
\end{enumerate}

In the following, for simplicity, the terms node and edge refer to their visual representation in a node-link layout, unless otherwise stated.

\begin{figure}[t]
\centering
\includegraphics[width=0.6\textwidth]{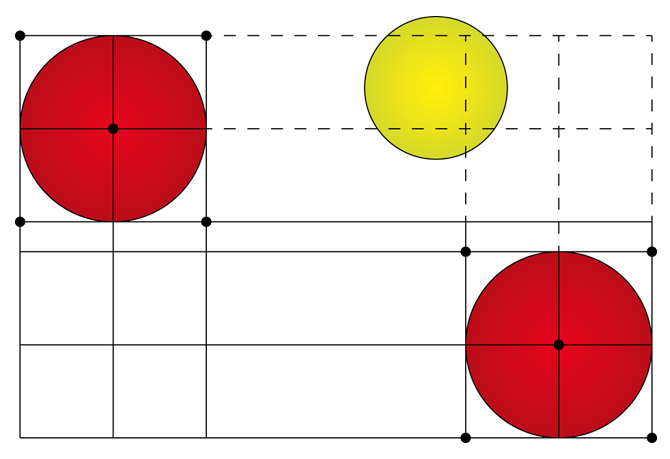}
\caption[Intersection test of orthogonal lines]{Intersection test of orthogonal lines between the connector points of the red nodes. Only the solid lines become edges of the OVG as the dashed ones intersect with the bounding box of the yellow node.}
\label{fig:intersection}
\end{figure}

\section{Computation of Potential Edge Routes}
The first step of the algorithm is to determine all potential orthogonal edge routes of the node-link representation.
For that, an auxiliary structure -- the so-called orthogonal visibility graph (OVG) -- is computed.
This structure describes all potential orthogonal edge routes and thus allows finding certain routes by applying search algorithms.
As this structure is based on the nodes, it must be updated upon every insert, update, or delete operation on the node set.

Nodes have a primary connector point in their center to which all incident edges connect, and four secondary connector points on the corners of their bounding boxes, through which edges can be routed.
The connector points are the nodes of the OVG.
The orthogonal edges of the OVG are calculated incrementally for pairs of nodes of the OVG. Let $v_1=(x_1,y_1)$ and $v_2=(x_2,y_2)$ be two nodes of the OVG, where $v_1$ belongs to the node $n_1$ of the node-link representation and $v_2$ to the node $n_2$.
Now the two orthogonal lines $e_1=(v_1,(x_1,y_2),v_2)$ and $e_2=(v_1,(x_2,y_1),v_2)$ are computed. If $x_1 = x_2$ or $y_1 = y_2$, then $e_1$ and $e_2$ are reduced to a single edge $e_3=(v_1,v_2)$.
For $e_1$, $e_2$, or $e_3$ it is tested whether they intersect with bounding boxes of existing nodes except $n_1$ and $n_2$. Non-intersecting lines become edges of the OVG. Figure \ref{fig:intersection} illustrates the computation of orthogonal lines for the two red nodes. In this case, only the solid lines become edges of the OVG. The dashed lines cannot be used because they intersect with the bounding box of an existing node (depicted in yellow).

In contrast to the OVG by \cite{Wybrow2009}, which contains only segments of potential orthogonal edge routes, the edges of the OVG used here can be clearly assigned to two nodes of the node-link representation (connector points have a reference to their nodes). This property is important for efficiently refreshing the OVG upon insert, update, and delete operations. 
If nodes are inserted, the dedicated nodes and edges of the OVG can be inserted easily as described above.
If nodes are deleted or updated, all affected edges of the OVG can be deleted in $O(|E_{OVG}|)$, whereas $E_{OVG}$ is the set of edges of the OVG.
If nodes are placed on existing edges of the OVG upon insert or update operations with the EditLens (verified with an intersection test), the respective edges of the OVG are marked as blocked and excluded from the following search. 
An example is depicted in Figure \ref{fig:intersection}. Assuming that the yellow node is placed after the red ones, the dashed edges are marked as blocked.
In case the blocking node is moved to a new position during an update operation, it is tested whether the blocked edges of the OVG are still blocked and released if necessary. If the blocking node is deleted, the blocked edges are released too.

The OVG used here and the OVG used by \cite{Wybrow2009} describe identical potential edge routes. However, the OVG used here requires more memory space.
The next step is to use the OVG to search for suitable edge routes.

\section{Search for Suitable Edge Routes}
With the OVG, the search for a suitable edge route between the primary connector points of two arbitrary nodes is easy.
If multiple edge routes must be found, the search must be performed multiple times.
For this purpose, a search algorithm for graphs can be used.
It is suggested to prefer fast, informed search algorithms like A-Star\footnote{A-Star is an informed search algorithm for graphs which is utilized to find the least-cost path between two given nodes of a graph. To enable a targeted and fast search, a heuristic cost function is used.} \cite{Hart1968}, which is also utilized in Wybrow et al.'s algorithm.
For finding edge routes that are short and have a low number of edge bends, the cost function of A-Star must be set up accordingly.
In case of large OVGs the search should be restricted temporally. If no route can be found in a certain amount of time (e.g., no route exists and the whole OVG must be traversed), a fall back solution must be used or the user must be prompted for further interaction.
After finding an edge route, its final representation will be computed.

\section{Resolution of Edge Overlaps}
\begin{figure}[t]
\centering
\includegraphics[width=0.5\textwidth]{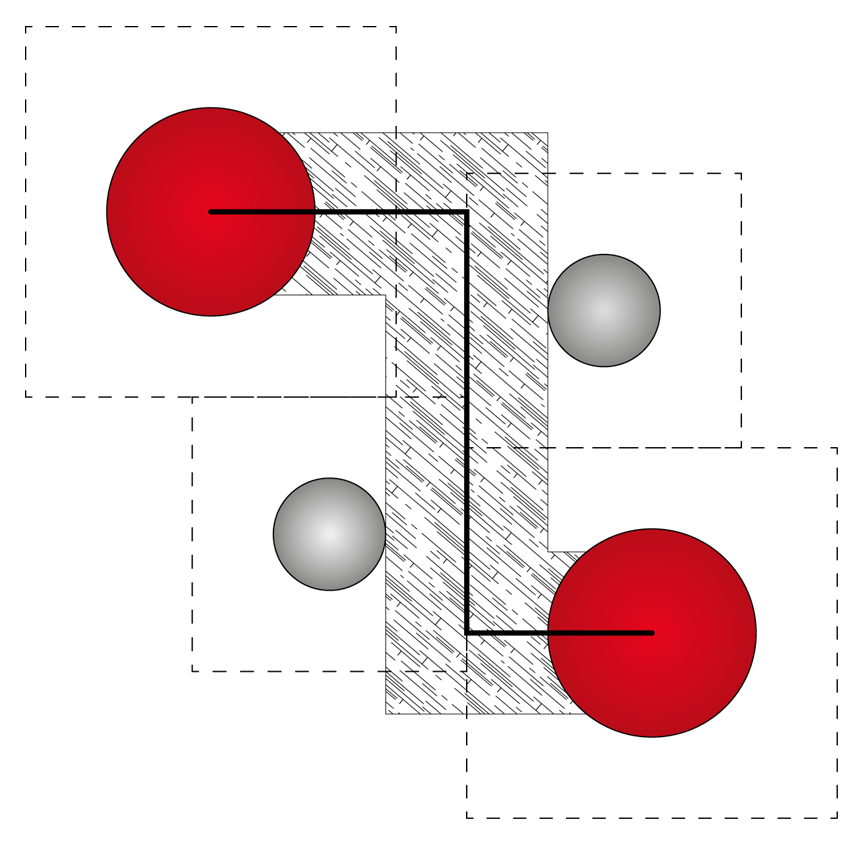}
\caption[Illustration of lanes between nodes used for edge routes]{Lanes between nodes used for edge routes (depicted as hatched area) resulting from enlarging the nodes bounding boxes by a safety gap $\delta$.}
\label{fig:lanes}
\end{figure}

The search for suitable edge routes on the OVG provides edge routes that might overlap partially. Using these routes in the node-link representation would introduce ambiguities and hinder tracing edges.
For that reason a method called nudging is used to resolve these overlaps.
The nudging by \cite{Wybrow2009} uses the available space to spread edges as much as possible and a global strategy to minimize their edge crossings.
With this nudging, good results can be achieved. However, it's runtime is too long for interactive frame rates as needed for the EditLens.
Thus, a simple local nudging is used here.
All edge routes provided by the search are shifted with an offset in $x$ and $y$ direction.
For avoiding crossings with existing nodes due to this shift, the bounding boxes of every node must be enlarged by a safety gap $\delta$. As this is already done when inserting nodes with the Editlens, no further adjustments are needed.
As illustrated in Figure \ref{fig:lanes}, this introduces lanes between all nodes with a thickness of at least $2\delta$. 
Within these lanes, the final edge routes are laid out.
The offset of the shift is computed as follows:
Let $r_{src}$ and and $r_{dest}$ be the radii of the source and the target node of an edge to be routed. Let $rand$ be a random number generator for the interval defined by its arguments and let $min$ be the minimum function.

\begin{equation*}
\begin{split}
\mbox{offset}_x = rand(0, min(r_{src}, r_{dest}, \delta-\mu))\\
\mbox{offset}_y = rand(0, min(r_{src}, r_{dest}, \delta-\mu))
\end{split}
\end{equation*}

$\mu$ is a parameter $>0$ that prevents edge routes from touching the border of nodes except for the source and target node.  
By shifting the edge routes with this offset, overlaps are often resolved and it is assured that the routes will connect to the circular nodes.

\begin{figure*}[h!]
	\centering
		\mbox{}\hfill
		\subfloat[]{\includegraphics[width=.9\textwidth]{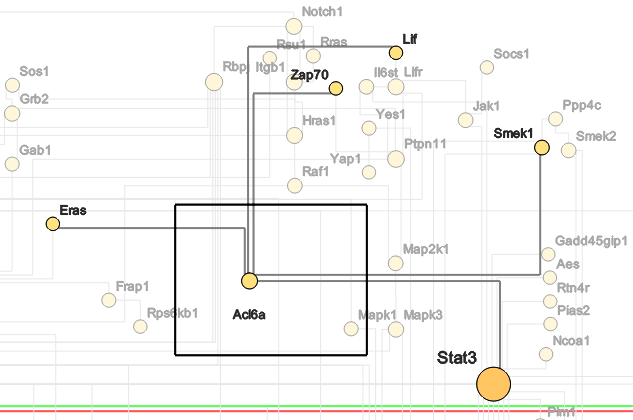}\label{fig:result1}}\hfill	\mbox{} \\
		\mbox{}\hfill
		\subfloat[]{\includegraphics[width=\textwidth]{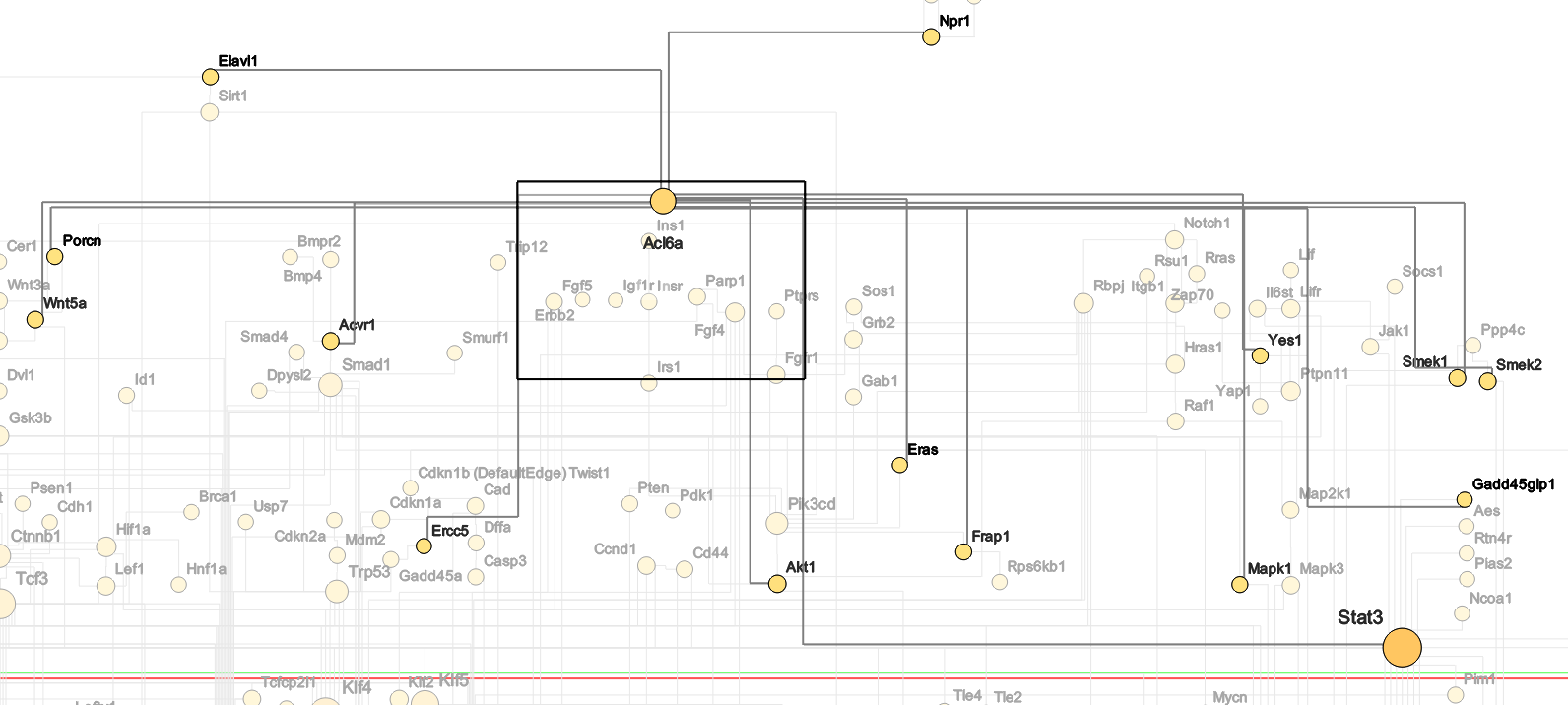}\label{fig:result2}}\hfill
		\mbox{}
		\caption[Two results obtained with the new edge routing algorithm]{The EditLens is used to interactively insert a node with \subref{fig:result1} 5 edges and \subref{fig:result2} 15 edges into the PluriNetwork. The inserted node is the one within the rectangular frame of the EditLens. The inserted edges are the ones shown fully saturated. The dimmed part of the layout is not affected by the ongoing edit operation.}
		\label{fig:results}
\end{figure*}

Figure \ref{fig:result1} presents a good result as obtained by this method during the insertion of a node with incident edges using the EditLens.
However, due to the randomness of the shift, overlaps can still appear in some cases. An example where this might happen are nodes with many incident edges.
Additionally, edge crossing introduced by this shift are not resolved. This can be seen in Figure \ref{fig:result2}.
Nevertheless, the user is free to explore different results with the EditLens until a good one is found where these problems do not (or only marginally) occur.

In \cite{Rohrschneider2009}, another local nudging strategy based on a topological ordering of a directed acyclic graph is used. It needs to be tested, whether this strategy is a suitable alternative.

\section{Discussion}

The proposed edge routing algorithm has been implemented in the EditLens prototype.
In order to achieve short runtimes, efficient data structures and parallel processing must be utilized.

For managing nodes, R-Trees are suitable. This hierarchical data structure offers an efficient intersection test needed for computing and updating the OVG. This intersection test can be realized for each segment of an edge route.
On the test computer (Intel Core i7-3770k at 4.66 GHz, 16 GB RAM), the intersection test of an edge segment with the nodes of the PluriNetwork takes only about 1-2 ms.
Another positive aspect of R-trees is that they provide important methods for inserting and deleting nodes, which are needed for incrementally constructing the graph layout.

The storage and the management of the OVG are difficult to handle. On the one hand, inserting and deleting its nodes and edges must be possible efficiently without recomputing the whole structure from scratch. For this purpose, linked lists can be used. On the other hand, the OVG should fit into the RAM to prevent slow access times due to outsourcing to the hard disk. This is already challenging with graphs of the PluriNetwork's size, as the number of edges of the OVG rises quadratically with its number of nodes. For this reason, only necessary information should be stored in the OVG.

For searching individual edge routes in the OVG, multiple computing threads can be used. This way, the search can be parallelized, which effectively reduces its runtime.

\begin{figure}[b!]
	\centering
	\includegraphics[width=\textwidth]{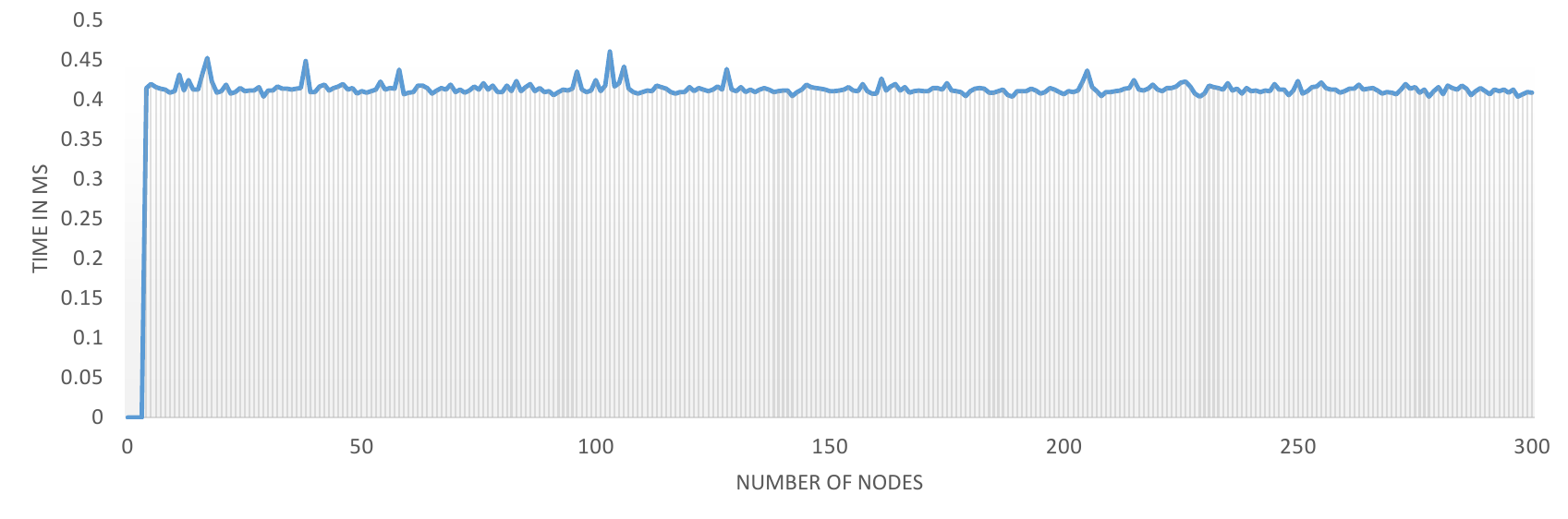}
	\caption{Results of a runtime performance test for computing orthogonal edge routes of a node-link layout with the edge routing algorithm proposed here.}
	\label{fig:chart2}
\end{figure}

To verify that the proposed edge routing algorithm is better or at least as good as the algorithm by \citep{Wybrow2009}, it was tested under the same conditions mentioned in the introduction. The average results of multiple runs are illustrated in Figure \ref{fig:chart2}. It shows that even with increasing number of nodes, the time needed to compute the two orthogonal edge routes remains constant at about 0.4 ms.
For comparison, recall the test results of Wybrow et al.'s algorithm that are depicted in Figure \ref{fig:chart}, where the runtime increases rapidly with increasing number of nodes.

Thanks to the short runtime of the new edge routing algorithm, using the EditLens for inserting nodes with 15 or more incident edges (as depicted in Figure \ref{fig:result2}) is usually possible at interactive frame rates.
However, there are also exceptional situations where the algorithms cannot compute an edge route in a given time limit (e.g., 500 ms).
The reasons for this are either non-existing edge routes or an extraordinary amount of nodes that must be circumvented.
With increasing number of incident edges upon node insertion, the different node placement strategies (e.g., edge length first strategy, see \cite{editlens}) can also be a limiting factor.

After developing the edge routing algorithm for the EditLens, \cite{Marriott2014} presented a novel orthogonal edge routing algorithm that is also significantly faster than Wybrow et al.'s algorithm. It remains to be tested whether their algorithm is a suitable alternative to be used for the EditLens.

\bibliographystyle{plainnat}

\bibliography{references}
\end{document}